\documentclass[aps,preprint,11pt]{revtex4}

\usepackage{amssymb,epsf}
\usepackage{epsfig}
\usepackage{epstopdf}
\usepackage{graphicx}
\usepackage{color}

\begin{document}

\title{\textbf{\Large Properties of the band gaps in one-dimensional ternary lossy photonic crystal containing double-negative materials}}
\author{A. Aghajamali $^{1,2}$\footnote{email address:
alireza.aghajamali@fsriau.ac.ir} M. Akbarimoosavi
$^{3}$\footnote{email address: maryammoosavim@yahoo.com}, and M.
Barati $^{2}$\footnote{email address: barati@susc.ac.ir}}

\affiliation{\small $^1$ Young Researchers Club, Science and Research Branch, Islamic Azad University, Fars, Iran\\
$^2$ Department of Physics, Science and Research Branch, Islamic Azad University, Fars, Iran\\
$^3$ Department of Physics, Shiraz University, Shiraz, Iran\\
}

\begin{abstract}
In this paper, theoretically, the characteristics matrix method is
employed to investigate and compare the properties of the band gaps
of the one-dimensional ternary and binary lossy photonic crystals
which are composed of double-negative and -positive materials. This
study shows that by varying the angle of incidence, the band gaps
for TM and TE waves behave differently in both ternary and binary
lossy structures. The results demonstrate that by increasing the
angle of incidence for the TE wave, the width and the depth of
zero-$\bar{n}$, zero-$\mu$, and Bragg gap increase in both ternary
and binary structures. On the other hand, the enhancement of the
angle of incidence for the TM wave, contributes to reduction of the
width and the depth of the zero-$\bar{n}$ and Bragg gaps, and they
finally disappear for incidence angles greater than $50^\circ$ and
$60^\circ$ for the binary structure, and $40^\circ$ and $45^\circ$
for the ternary structures, respectively. In addition, the details
of the edges of the band gaps variations as a function of incidence
angle for both structures are studied.
\end{abstract}\

\maketitle

\section{Introduction}

Photonic crystals (PCs) are artificial dielectric or metallic
structures in which the refractive index changes periodically. The
periodic structure of the PCs contributes to photonic band gap
(forbidden range of frequencies). Interference of the Bragg
scatering is considered as a cause of this phenomena. PCs have
attracted much interest due to their novel electromagnetic waves
characteristics and important scientific and engineering
applications and have received attentions by many researches in
recent decades \cite{re1,re2,re3,re4,re5,re6,re7,re8}. With
possibility of producing metamaterials, PCs containing
metamaterials, namely ``Metamaterial Photonic Crystals'' (MPCs) have
been made. The technological relevance of the MPCs is of great
importance. These structures have attracted considerable attentiom
for their various applications. Designing dielectric mirrors,
effective waveguides, filters, and perfect lenses are clear cut
examples. The metamaterials are utalized to design binary and
ternary MPCs with layers of Double Negative (DNG) and Double
Positive (DPS) materials. In these structures, various band gaps can
be seen in the transmission spectrum. When the average refractive
index of the MPC is equal to zero, the band gap is called
zero-$\bar{n}$ gap \cite{re9}. The properties of the zero-$\bar{n}$
gap of one-dimensional (1D) binary MPC that composed of the two
layers of DNG and DPS materials were investigated in recent years
\cite{re9,re10,re11,re12,re13,re14,re15,re16}. In 1D binary PMC, in
addition to the zero-$\bar{n}$ and Bragg gaps, we also have
zero-$\mu$ and zero-$\varepsilon$ gaps for TE and TM waves
respectively, which appear for the oblique incidence angles. The
results have been also reported by some authors
\cite{re16,re17,re18}.

The novel idea of this research is to include the loss factor which
is an unavoidable consequence of Double Negative materials to
investigate optical properties of 1D binary and ternary MPCs. In
this regard the transmission of electromagnetic waves through a 1D
ternary lossy PC consisting of layers with double-negative and
-positive materials is studied. The effects of the incidence angle
and polarization on the characteristics of the zero-$\bar{n}$,
zero-$\mu$, zero-$\varepsilon$, and Bragg gaps are investigated.
Furthermore, the results are compared with the similar results in
the 1D binary lossy PC.

The paper is organized as follows: the MPC structures design, the
permittivity and permeability of the DNG layer, and theoretical
formulation (characteristic matrix method) are described in Section
2, the numerical results and discussions are presented in Section 3,
and the paper is concluded in Section 4.

\section{Structures design and characteristic matrix method}

The 1D MPC structure under study which is located in air is
constituted by alternative layers of DNG and DPS materials, where
the DNG material is dispersive and dissipative. We consider the 1D
binary MPC with periodic structure of $(AB)^N$ and the 1D ternary
MPC with periodic structure of $(ABC)^N$, where \emph{A}, \emph{B},
and \emph{C} denote three different materials. \emph{N} is number of
the lattice period and $d_i$, $\varepsilon_i$, and $\mu_i$
$(i=A,B,C)$ are thickness, permittivity and permeability of the
layers, respectively.

The permittivity and permeability of the DNG (layer \emph{A}) in the
microwave region are complex and are given as \cite{re9},
\begin{equation}
\label{nineeq}
      \varepsilon_{A}(f) =1+\frac{5^2}{0.9^2-f^2-i\gamma f}+\frac{10^2}{11.5^2-f^2-i\gamma f}
\end{equation}
\begin{equation}
\label{teneq}
      \mu_{A}(f) =1+\frac{3^2}{0.902^2-f^2-i\gamma f}
\end{equation}
where \emph{f} and $\gamma$ are frequency and damping frequency,
respectively, given in GHz. Various details of the real parts of the
permittivity and permeability of layer \emph{A}, versus frequency
have been discussed in our previous work \cite{re10}.

The calculation is based on the characteristic matrix method
\cite{re19}, which is most effective to analyze the transmission
properties of PCs. The characteristic matrix $M[d]$ for TE wave at
incidence angle from vacuum to a 1D PC structure is given by
\cite{re19}:
\begin{equation}
M_i= \left[
\begin{array}{cc}
\cos \gamma_{i}  &  \frac{-i}{p_{i}} \ \sin \gamma_{i}\\
-i \ p_{i} \ \sin \gamma_{i}  &  \cos \gamma_{i}
\end{array}\right]
\end{equation}
where $k=2$ and $3$ for binary and ternary PCs, respectively.
$\gamma_{i}=(\omega /c) \: n_{i} d_{i} \cos\theta_{i} $, \emph{c} is
speed of light in vacuum, $\theta_{i}$ is the angle of refraction
inside the layer with refractive index $n_i$ and $p_i$ is given as:
\begin{equation}
p_{i}=\sqrt{\frac{\varepsilon_{i}}{\mu_{i}}}\:\cos\theta_{i}
\end{equation}
where
\begin{equation}
\cos\theta_{i}=\sqrt{1-\frac{n_{0}^2\:\sin^2\theta_{0}}{{n_{i}^2}}}
\end{equation}

The characteristic matrix for \emph{N} period structure is therefore
$[M(d)]^N$. The transmission coefficient of the multilayer is given
by:
\begin{equation}
t=\frac{2\ p_{0}}{(m_{11}+m_{12}\ p_{s})\ p_{0}+(m_{21}+m_{22}\
p_{s})}
\end{equation}
where $m_{ij} (i,j=1,2)$ are the matrix elements of $[M(d)]^N$, and
\begin{equation}
p_{0}=n_{0}\ \cos\theta_{0} , p_{s}=n_{s}\ \cos\theta_{s}
\end{equation}

The transmissivity of the multilayer is given by:
\begin{equation}
T=\frac{p_{s}}{p_{0}} |t|^2 .
\end{equation}

The above formulations can be applied for TM wave by simple
replacements of $p_i$, $p_0$, and $p_s$ as follows:
\begin{equation}
p_{i}=\sqrt{\frac{\mu_{i}}{\varepsilon_{i}}}\cos\theta_{i}
\end{equation}
\begin{equation}
p_{0}=\frac{\cos\theta_{0}}{n_{0}} ,
p_{s}=\frac{\cos\theta_{s}}{n_{s}}.
\end{equation}

\section{Numerical results and discussion}

Based on the theoretical model described on the previous section,
the transmission spectrum of the presented lossy MPC structures was
calculated. In the study of the 1D binary MPC, consisting of DNG
(layer \emph{A}) and DPS (layer \emph{B}) materials, equations (1)
and (2) are used for the permittivity, $\varepsilon_A$, and the
permeability, $\mu_A$, of layer \emph{A}. Layer \emph{B} is assumed
to be a vacuum layer with $\varepsilon_B=\mu_B=1$ $(n_B=1)$. The
thickness of layers \emph{A} and \emph{B} are chosen as $d_A=6$mm
and $d_B=12$mm, respectively and the total number of lattice periods
is set as $N=16$ \cite{re10}.

The transmission spectra of TE and TM polarized waves for the binary
structure at various angles of incidence and for $\gamma =
0.2\times10^{-3}$ GHz are shown in Figs. 1(a) and 1(b),
respectively. For oblique incidence the zero-$\mu$ and
zero-$\varepsilon$ gaps appear in the transmission spectrum for TE
and TM waves, respectively, as reported in \cite{re17,re18}. The
gaps appear at the frequencies where the sign of the permeability or
permittivity of DNG materials changes. As it is seen in Fig.1 the
width of the gaps increases as the incidence angle increases while
the left edge of the gaps remain nearly unchanged. In addition, for
TE waves the width and the depth of the zero-$\bar{n}$ and the Bragg
gap increase as the incidence angle increases, as reported in
\cite{re9,re10,re11,re12,re13,re14}. On the contrary, for TM waves
the width and the depth decrease as the incidence angle increases
(Fig. 1(b)). Moreover, the figures show that the central frequency
of the zero-$\bar{n}$, as reported in our previous work \cite{re10},
and the Bragg gap shift to the higher frequencies as the angle of
incidence increases.

In the next part, the band gaps of the 1D ternary MPC are
investigated. The binary structure used before, is modified by
introducing a third layer of $SiO_2$ (layer \emph{C}) with
refractive index of $n_C=1.46$ and thickness of $d_C=6$mm in each
lattice period. The other parameters are kept the same as in the
binary structure. The transmission spectra of TE and TM polarized
waves for the ternary structure at various angles of incidence and
for $\gamma = 0.2\times10^{-3}$ GHz are shown in Figs. 2(a) and
2(b), respectively. As it is seen, the width of the
zero-$\varepsilon$ and zero-$\mu$ gaps increases by increasing the
incidence angle. Although, for TE wave the width and the depth of
the zero-$\bar{n}$ and the Bragg gaps increases as the incidence
angles increases. For TM waves the behaviour is completely
different, such that the width and the depth of the band gaps
decrease when the angle of incidence increases. However, the central
frequency of the zero-$\bar{n}$ and Bragg gaps behave the same as in
the binary structure where they shift to the higher frequencies as
the angle of incidence increases.

In this part the band gaps in binary and ternary structures are
compared. As it is seen from, Figs. 1 and 2, the width of the band
gaps decreases, when another dielectric layer added to the lattice
period of binary structure. Furthermore, the zero-$\bar{n}$ and
Bragg gaps frequencies of the ternary structure are lower the
corresponding frequencies in the binary structures; while the
zero-$\varepsilon$ and zero-$\mu$ gaps appears in the same
frequencies in both structures.

The lower ($f_L$) and higher ($f_H$) frequencies of the
zero-$\bar{n}$ gap in the binary and ternary structures, as a
function of incidence angle for TE and TM waves are shown in Figs.
3(a) and 3(b), respectively. As it is clearly seen, the lower
frequency of the gap for the ternary structure is smaller than that
in the binary structure for both TE and TM modes for different
incidence angles. It is also interesting to note that the width of
the gap in the ternary structure for TE wave is sensitive to the
angle of incidence and increases as the incidence angle increases
while the left edge of the band gap is nearly invariant. Moreover,
the width of the zero-$\bar{n}$ gap of both binary and ternary
structures is very sensitive to the incidence angle for TM wave, and
the gap disappears for incidence angles greater than $50^\circ$ in
the binary structure as reported in [10,12]. The gap disappears for
incidence angles greater than $40^\circ$ for the ternary structure.

The lower and the higher frequencies of the zero-$\mu$ and
zero-$\varepsilon$ gaps as a function of incidence angle, for both
binary and ternary structures are shown in Figs. 4(a) and 4(b),
respectively. The width of the zero-$\mu$ and zero-$\varepsilon$
gaps increases faster for the binary structure than the ternary
ones. As it is clearly seen, the gaps appears at the same frequency
for both binary and ternary structures; in addition, the figures
indicate that the upper edge of the gaps, so the width of the gaps,
are very sensitive to the incidence angle while the lower edge of
the gaps are nearly insensitive.

The $f_L$ and $f_H$ frequencies of the Bragg gap of the binary and
ternary structures, as a function of incidence angle for TE and TM
waves are shown in Figs. 5(a) and 5(b), respectively. It is seen
that the Bragg gap like the zero-$\bar{n}$ gap appears in the lower
frequencies for the ternary structure. Moreover, $f_L$ and $f_H$
shift to the higher frequencies as the incidence angle increases,
and as it is seen from Fig. 5(b), the gap for TM polarized wave
disappears for incidence angles greater than $60^\circ$ for the
binary and $45^\circ$ for the ternary structures.

\section{Conclusion}

The numerical results show that including the loss factor in the
permittivity and permeability of the DNG layer the changes in
transmission spectrum of TE wave in 1D ternary and binary PMCs
behave similarly for different incidence angles. The zero-$\bar{n}$
and zero-$\mu$, and the Bragg gaps become wider and dipper as the
angle of incidence increases. In addition, by enhancing the angle of
incidence, the lower frequencies of the zero-$\bar{n}$ and
zero-$\mu$ gaps are nearly invariant in both ternary and binary
PMCs, but the Bragg gap increases. Moreover, the upper edge of all
three band gaps increases. Such behaviors do not clearly observed in
the zero-$\bar{n}$ gap. The transmission spectrum for TM polarized
wave is rather different. It is found that in both structures, as
the angle of incidence increases, the width and the depth of the
zero-$\varepsilon$ gap increase, while for the zero-$\bar{n}$ gap
and for the angles of incidence greater than $40^\circ$ for the
ternary and $50^\circ$ for the binary structures the zero-$\bar{n}$
gap disappears. Similar behavior is observed for the Bragg gap for
the angles of incidence greater than $45^\circ$ and $60^\circ$ for
binary and ternary structures, respectively. For TM waves, the lower
frequency of the zero-$\bar{n}$ and Bragg gaps increases, but the
zero-$\varepsilon$ gap decreases as the angle of incidence increases
in both binary and ternary structures. The higher edge of the
zero-$\varepsilon$ and Bragg gaps increases while it decreases for
the zero-$\bar{n}$ gap. It is interesting to note that the
difference between the lower and higher frequencies of all the gaps
for TM wave in the binary PMC, is larger than that in the ternary
PMC as all the figures show. To recapitulate what was said before,
i- the width of the band gaps decreases as a dielectric layer is
added to the binary structure. ii- the zero-$\bar{n}$ and Bragg gaps
appear at the lower frequencies in the ternary structure. iii- the
zero-$\varepsilon$ and zero-$\mu$ gaps appear at the same
frequencies in both structures. Finally, the results of the present
study could be employed in designing new edge filters, waveguides,
and other optical devises in microwave engineering.


\newpage
\thispagestyle{empty}
\begin{figure}[tbp]
\epsfxsize=7cm \centerline{\includegraphics [width=15cm]
{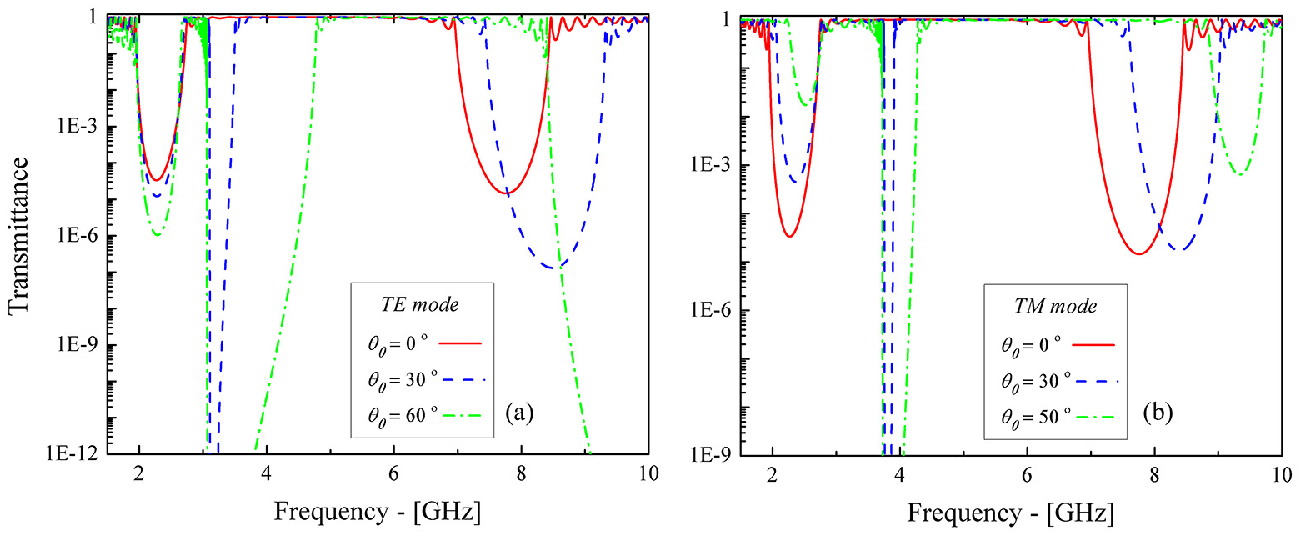}} \caption{Transmission spectra for the 1D binary MPC
structure, for different angles of incidence, with $\gamma =
0.2\times10^{-3}$ GHz; (a) TE (b) TM waves.}
\end{figure}

\begin{figure}[tbp]
\epsfxsize=7cm \centerline{\includegraphics [width=15cm]
{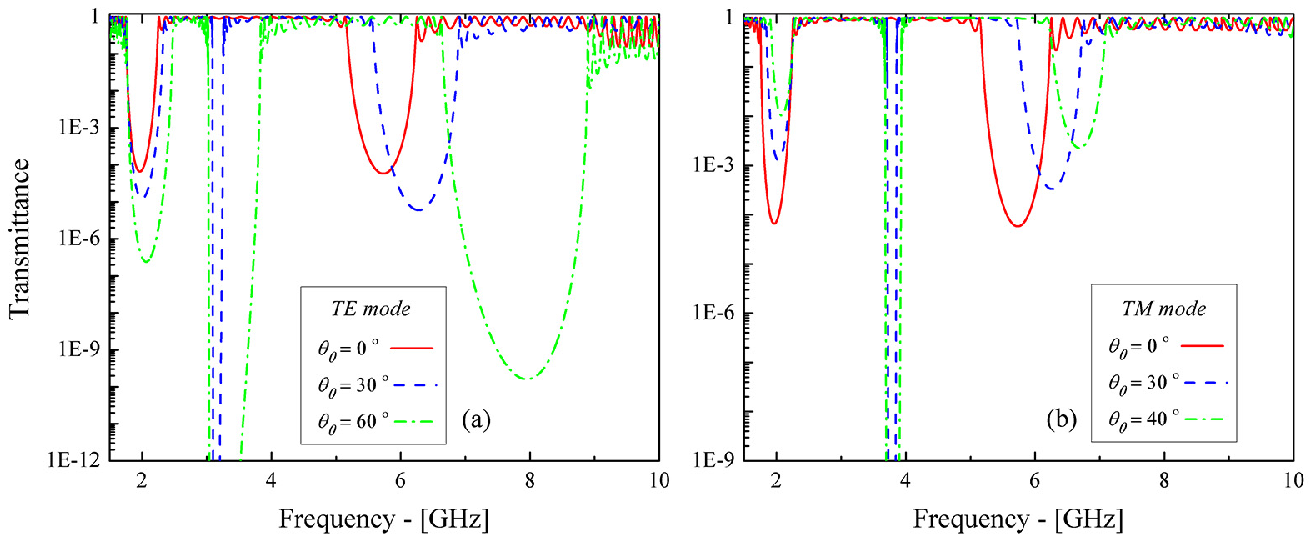}} \caption{Transmission spectra for the 1D ternary MPC
structure, for different angles of incidence, with $\gamma =
0.2\times10^{-3}$ GHz; (a) TE (b) TM waves.}
\end{figure}

\begin{figure}[tbp]
\epsfxsize=7cm \centerline{\includegraphics [width=15cm]
{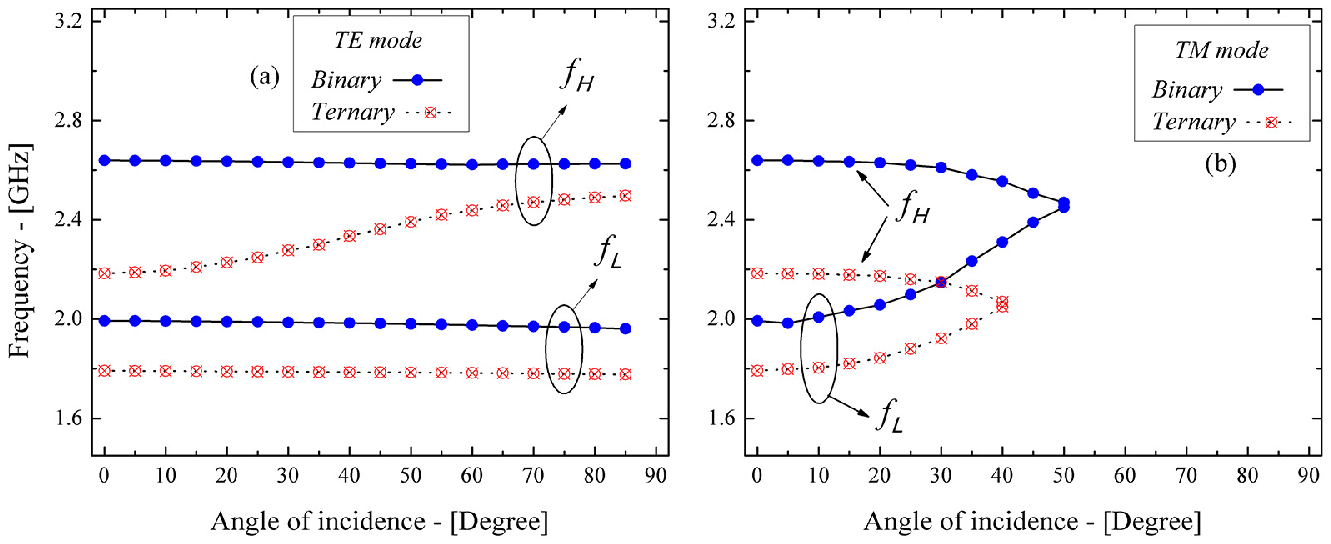}} \caption{The lower and higher frequencies of the
zero-$\bar{n}$ gap as a function of angle of incidence for both
binary and ternary structures; (a) TE and (b) TM waves.}
\end{figure}

\begin{figure}[tbp]
\epsfxsize=7cm \centerline{\includegraphics [width=15cm]
{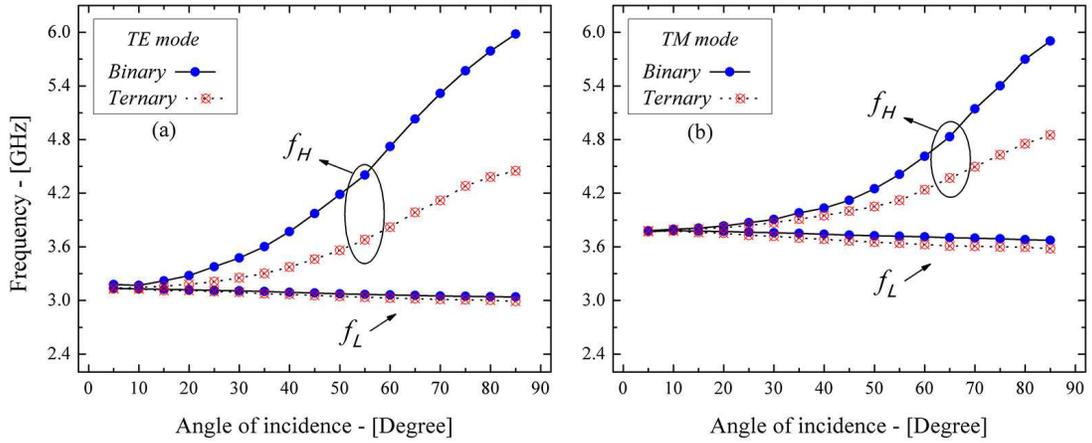}} \caption{The lower and higher frequencies of the (a)
zero-$\mu$ and (b) zero-$\varepsilon$ gaps as a function of angle of
incidence for both binary and ternary structures.}
\end{figure}

\begin{figure}[tbp]
\epsfxsize=7cm \centerline{\includegraphics [width=15cm]
{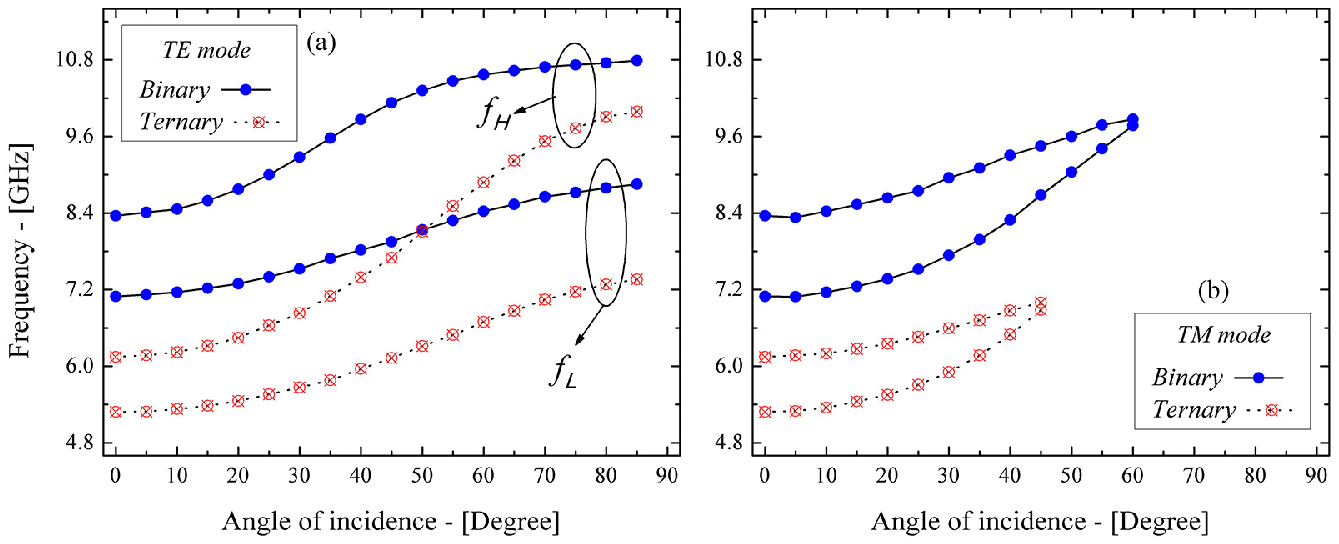}} \caption{The lower and higher frequencies of the Bragg
gap as a function of angle of incidence for both binary and ternary
structures; (a) TE and (b) TM waves.}
\end{figure}


\end{document}